\documentclass[aps,prl,twocolumn,superscriptaddress,letterpaper]{revtex4}

\usepackage{graphicx}
\usepackage{dcolumn}
\usepackage{bm}
\usepackage{amsmath, amssymb}%
\usepackage{epstopdf}
\usepackage{color}
\usepackage[titletoc,toc,title]{appendix}
\usepackage{lipsum}

\usepackage{hhline}
\usepackage{multirow}

\usepackage{float}
\floatstyle{plaintop}
\restylefloat{table}

\usepackage[]{changes} 

\begin{document}

\title{Short-range correlated pair formation and nuclear shell structure}
  

\newcommand*{\NOWCUA}{Catholic University of America, Washington, DC
  20064, USA}
\newcommand*{\NOWUTK}{University of Tennessee, Knoxville, TN 37919, USA}
\newcommand*{\NOWFIU}{Florida International University, Miami, FL 33199, USA}
\newcommand*{\JLab}{Thomas Jefferson National Accelerator Facility, Newport News, Virginia 23606, USA}
\newcommand*{\JLabindex}{1}
\affiliation{\JLab}

\newcommand*{\UTK}{University of Tennessee, Knoxville, Tennessee 37919, USA}
\newcommand*{\UTKindex}{2}
\affiliation{\UTK}

\newcommand*{\ODU}{Old Dominion University, Norfolk, Virginia 23529, USA}
\newcommand*{\ODUindex}{3}
\affiliation{\ODU}

\newcommand*{\CUA}{The Catholic University of America, Washington, DC 20064, USA}
\newcommand*{\CUAindex}{4}
\affiliation{\CUA}

\newcommand*{\FIU}{Florida International University, Miami, Florida 33199, USA}
\newcommand*{\FIUindex}{5}
\affiliation{\FIU}

\newcommand*{\MIT}{Massachusetts Institute of Technology, Cambridge, Massachusetts 02139, USA}
\newcommand*{\MITindex}{6}
\affiliation{\MIT}

\newcommand*{\TAU}{School of Physics and Astronomy, Tel Aviv University, Tel Aviv 69978, Israel}
\newcommand*{\TAUindex}{7}
\affiliation{\TAU}

\newcommand*{\LBNL}{Lawrence Berkeley National Laboratory, Berkeley, California 94720, USA}
\newcommand*{\LBNLindex}{8}
\affiliation{\LBNL}

\newcommand*{\UW}{University of Washington, Seattle, Washington 98195, USA}
\newcommand*{\UWindex}{9}
\affiliation{\UW}

\newcommand*{\GWU}{The George Washington University, Washington, DC 20052, USA}
\newcommand*{\GWUindex}{10}
\affiliation{\GWU}

\newcommand*{\WM}{The College of William \& Mary, Williamsburg, Virginia 23185, USA}
\newcommand*{\WMindex}{11}
\affiliation{\WM}

\newcommand*{\NRCN}{Nuclear Research Center, Negev, Beer-Sheva, Israel 84190}
\newcommand*{\NRCNindex}{12}
\affiliation{\NRCN}

\newcommand*{\MSU}{Mississippi State University, Mississippi State, Mississippi 39762, USA}
\newcommand*{\MSUindex}{13}
\affiliation{\MSU}

\newcommand*{\UVA}{University of Virginia, Charlottesville, Virginia 22903, USA}
\newcommand*{\UVAindex}{14}
\affiliation{\UVA}

\newcommand*{\UConn}{University of Connecticut, Storrs, Connecticut 06269, USA}
\newcommand*{\UConnindex}{15}
\affiliation{\UConn}

\newcommand*{\HU}{Hampton University, Hampton, Virginia 23669, USA}
\newcommand*{\HUindex}{16}
\affiliation{\HU}

\author{D. Nguyen}
\thanks{Equal Contribution}
\affiliation{\JLab}\affiliation{\UTK}
\thanks{Equal Contribution}
\author{C. Yero}
\thanks{Equal Contribution}
\affiliation{\ODU}\affiliation{\CUA}
\author{H. Szumila-Vance}
\affiliation{\JLab}\affiliation{\FIU}
\author{F. Hauenstein}
\affiliation{\JLab}
\author{N. Swan}
\affiliation{\ODU}
\author{L.B. Weinstein}\email[Contact Author \ ]{weinstein@odu.edu}
\affiliation{\ODU}
\author{J.~Kahlbow}\affiliation{\MIT}\affiliation{\TAU}\affiliation{\LBNL}
\author{A. Schmidt}
\affiliation{\GWU}
\author{E. Piasetzky}
\affiliation{\TAU}
\author{O. Hen}
\affiliation{\MIT}

\author{C.~Ayerbe~Gayoso}\affiliation{\WM}\affiliation{\ODU}
\author{E.~Cohen}\affiliation{\TAU}\affiliation{\NRCN}
\author{P.~Datta}\affiliation{\UConn}
\author{A.~Denniston}\affiliation{\MIT}
\author{B.R.~Devkota}\affiliation{\MSU}
\author{M.~Diefenthaler}\affiliation{\JLab}
\author{C.~Fogler}\affiliation{\ODU}
\author{B.R.~Gamage}\affiliation{\JLab}
\author{D.~Higinbotham}\affiliation{\JLab}
\author{I.~Korover}\affiliation{\TAU}
\author{C.~Morean}\affiliation{\UTK}
\author{M.~Nycz}\affiliation{\UVA}
\author{M.~Satnik}\affiliation{\WM}
\author{S.~Seeds}\affiliation{\UConn}
\author{P.~Sharp}\affiliation{\GWU}
\author{I.~Shlush}\affiliation{\TAU}
\author{M.~Suresh}\affiliation{\HU}
\author{A.S.~Tadepalli}\affiliation{\JLab}
\author{R.~Wagner}\affiliation{\TAU}
\author{E. W.~Wertz}\affiliation{\WM}

\collaboration{The Hall C Collaboration}
\noaffiliation

\date{\today}


\begin{abstract}
Short-range correlated (SRC) nucleon pairs, caused by brief, high-momentum interactions between two nucleons, are dominated by neutron-proton pairs with large relative and smaller center-of-mass momenta. However, which nucleons form pairs is not fully understood.  Measurements of $^{40,48}$Ca and $^{54}$Fe showed that intra-shell pairing was much stronger than inter-shell pairing. 
We measured high-missing-momentum protons knocked out in electron scattering from selected nuclei with a range of  shell configurations,  $A$, and $N/Z$, including $^9$Be, $^{10,11}$B, $^{12}$C, $^{40,48}$Ca, $^{54}$Fe, and $^{197}$Au, focusing on  the four light  nuclei. 
Unexpectedly, we found that adding one proton to  $^9$Be (to make $^{10}$B) increased the number of correlated protons far more than adding one neutron to $^{10}$B or adding one proton to $^{11}$B, perhaps due to the large spin ($J=3$) of $^{10}$B or the unusual structure of $^9$Be.  In addition, the increase in pairing probability from $^9$Be to $^{12}$C and from $^{40}$Ca to $^{54}$Fe was much steeper than the increase from $^9$Be to $^{197}$Au.  This shows  the importance of  low-momentum nuclear  structure on the probability of high-momentum nucleon pairing. 
  
\end{abstract}

\maketitle 
The independent-particle shell model provides a first-order description of nuclei in which nucleons occupy  mean-field orbitals. However,  valence orbitals contain only about two-thirds   the expected number of protons~\cite{lapikas97}.  Instead, many of these ``missing'' nucleons belong to temporary high-momentum short-range correlated (SRC) nucleon-nucleon ($NN$) pairs~\cite{subedi08}. These nucleons account for most of the kinetic energy of the nucleons and have been linked to the modification of bound-nucleon structure~\cite{Hen:2016kwk} and to neutron star properties~\cite{Gautam:2024qam}, neutrino-less double beta decay~\cite{Deppisch:2020ztt,Kortelainen:2007rh} and accelerator-based neutrino-oscillation experiments~\cite{electronsforneutrinos:2020tbf}.

Theoretical studies predict universal short-range factorization and interaction-independent SRC scaling, suggesting that SRC abundances reflect long-range nuclear structure rather than short-distance dynamics~\cite{Vanhalst:2011es,Chen:2016bde,Lynn:2019vwp,Cruz-Torres:2019fum}. However, quantitative predictions remain limited for heavier nuclei.

Experiments have established that nearly all high-momentum nucleons belong to SRC pairs, and these are predominantly proton–neutron pairs with large relative and smaller center-of-mass momentum \cite{Hen:2016kwk,Piasetzky:2022skb,subedi08,duer18,Cohen:2018gzh,hen14,CLAS:2020rue,CLAS:2022odn}.

However, there is very limited information about which nucleons form correlated pairs and whether pairing is predominantly determined by nuclear mass $A$,  density,  asymmetry (the neutron-to-proton ratio $N/Z$), or  shell structure.

Inclusive $A(e,e')$ electron-scattering measurements showed that the fraction of nucleons in correlated pairs increases by about 10\% from C to Pb, consistent with nuclear saturation~\cite{egiyan02,egiyan06,fomin12,Schmookler:2019nvf}. An inclusive measurement of $^{48}$Ca and $^{40}$Ca  showed that increasing the number of neutrons by 40\% in $^{48}$Ca only increased the number of correlated nucleons by 17\%~\cite{Nguyen:2020mgo}. However, inclusive measurements suffer from interpretation uncertainties~\cite{Weiss:2020mns} and provide no information on which nucleons form pairs.

Measurements of  $A(e,e'p)$ and $A(e,e'n)$ cross-section ratios on C, Al, Fe and Pb showed that the fraction of SRC protons (as measured by the  ratio of high-initial-momentum to low-initial-momentum protons) increased by a factor of 1.5 from carbon to lead and the same ratio for neutrons was constant~\cite{duer18}.  However, the limited selection of target nuclei precluded specifically attributing the increased fraction of SRC protons to  the effects of nuclear mass ($A$) or nuclear asymmetry ($N/Z$).

A recent measurement of $(e,e'p)$ proton-knockout  from SRC pairs in  $^{40}$Ca, $^{48}$Ca, and $^{54}$Fe~\cite{CaFeNature} showed that adding eight $1f_{7/2}$ neutrons to $^{40}$Ca only increased the number of high-initial-momentum protons by $10\pm 2\%$, but that adding six $1f_{7/2}$ protons increased the  number of high-initial-momentum protons by a surprising 50\% ($1.49\pm0.03$). These results challenged previous expectation that SRC pair formation is governed by global properties such as mass $A$, or neutron-to-proton asymmetry $N/Z$ and indicated that \textit{intra}-shell SRC pairing is far more favorable than \textit{inter}-shell pairing.

In addition to the CaFe nuclei results reported above, the same experiment  measured  electron-scattering proton-knockout cross-section ratios from SRC pairs in
 a range of nuclei spanning mass $A$, asymmetry $N/Z$, and shell occupancy:
$^9$Be, $^{10}$B, $^{11}$B, $^{12}$C, $^{40}$Ca, $^{48}$Ca, $^{54}$Fe, and $^{197}$Au.



This experiment ran in 2022--2023 in the Thomas Jefferson National Accelerator Facility's (Jefferson Lab) Hall C, using a 10--60 $\mu$A,  10.5~GeV electron beam. The scattered electrons and protons were detected in the SHMS and HMS spectrometers, respectively~\cite{ShmsNim,HmsNim,HallC:2022qlb}. Both spectrometers were equipped with  pairs of drift chambers for particle tracking, scintillator hodoscope planes for event triggering and timing, and an electromagnetic calorimeter for electron identification. The boron targets were B$_4$C, and the $^{48}$Ca target was 90\% $^{48}$Ca and 10\% $^{40}$Ca, by number of atoms.

The SHMS was positioned at a central angle $\theta_e= 8.3^{\circ}$ and momentum  $p_e = 8.55$ GeV/$c$.  The average final electron momentum after cuts was 9.72 GeV/$c$, giving a central four-momentum transfer $Q^2 = \vec{q}\thinspace^2 - \omega^2= 1.97$ GeV$^2$/$c^2$ where the three-momentum transfer is $\vec{q} = \vec{p}_e - \vec{p}_e{}^{'}$, 
energy transfer $\omega = E_{\text{beam}} - E^{'}_e$ and $x_B = Q^2/2m_p\omega$ ($m_p$ is the proton mass). The HMS was set at central angle $\theta_p = 66.4^{\circ}$ and momentum  $p_p = 1.325$ GeV/$c$ covering a range of missing momentum $250 \leq p_\mathrm{miss} \leq 700$ MeV/$c$ where $\vec p_\mathrm{miss} = \vec p_p - \vec q$. In the plane wave impulse approximation (PWIA), where  the knocked-out proton leaves the nucleus without  rescattering, the initial momentum of the struck proton is $p_i = p_\mathrm{miss}$, and the separation energy of the proton is $E_i = E_{miss} = \omega - T_p$, where $T_p$ is the kinetic energy of the detected proton.

We identified electrons by requiring that the ratio of deposited energy in the calorimeter to the particle momentum was $0.8<\frac{E_{\text{cal}}}{|p_{e'}|}<1.3$, and we identified coincident  $(e,e'p)$ events by applying a $\pm 2$ ns cut around the $e'$ and $p$ coincidence time peak. To avoid spectrometer edge effects, we applied  geometrical collimator cuts.

We validated the performance of the spectrometers by comparing  the measured hydrogen elastic electron scattering cross section to the world data.  These  cross sections agreed to within 5\%.  Any spectrometer inefficiencies cancel in our cross section ratios.

To select protons knocked out from SRC pairs,  we required $375 \le |p_\mathrm{miss}| \le 700$ MeV/$c$, where 375 MeV/$c$ corresponds approximately to the onset of $NN$ SRC pair dominance \cite{CLAS:2022odn}.  To isolate quasi-elastic scattering events from SRC pairs, we  required  $Q^{2} \ge 1.8$ (GeV/$c$)$^2$ to  suppress the non-quasielastic contribution from meson exchange currents~\cite{Arrington:2011xs}. To reduce inelastic scattering contributions, such as from nucleonic excitations, we required $x_{B}\ge 1.2$~\cite{Arrington:2011xs}.  Finally, to minimize the effect of low $p_\mathrm{miss}$ protons rescattering and appearing in our data sample at higher $p_\mathrm{miss}$, we required  that the angle between the reconstructed recoil momentum ($\vec{p}_{\text{recoil}} = - \vec{p}_{\text{miss}}$) and $\vec{q}$ be small: $\theta_{rq} < 40^{\circ}$~\cite{CLAS:2007tee,Yero:2020cbq}. See Supplemental Materials for plots of these quantities.
 

To extract the normalized yields,  we first corrected the measured number of events run-by-run by  the product of the spectrometer tracking efficiencies and electronic live time.  This factor was typically $\approx 97\%$. We then normalized these yields by dividing by the integrated charge, $Q_A$ and target areal proton density $t_A$.  We  determined the boron normalized yields by subtracting the scaled measured carbon yields from the B$_4$C target yields.  We corrected the  ``$^{48}$Ca-target" normalized yield by subtracting the scaled measured $^{40}$Ca yield. We further corrected the $^{40,48}$Ca yields for oil contamination (see Supplemental Information for details).

The calculated cross section ratios of each nucleus $A$ to carbon are further corrected for radiative ($R_A/R_C$) and transparency ($T_A/T_C$) effects: 
\begin{equation}
    \sigma_{A/C}= \frac{\sigma_A}{\sigma_C} = \frac{N_A /(Q_A\epsilon_A t_A T_A R_A)}{N_C /(Q_C\epsilon_C t_C T_C R_C)} ,
    \label{eq:ratio}
\end{equation}
where $N_A$ are the yields corrected for target impurities and
contamination.  The radiative correction factors, $R_A$, correct for the effects of electron radiation in the electric field of the target nucleus.  They were
determined from the ratio of the radiated-to-unradiated PWIA cross
section calculated using the Hall C SIMC simulation~\cite{simcref},
which incorporated the Benhar Spectral function~\cite{BENHAR1994493}
for $^{12}$C, $^{56}$Fe and $^{197}$Au and was integrated over our
experimental kinematics. The correction factors were $0.741$, $0.734$,
and $0.603$ for $^{12}$C, $^{56}$Fe and $^{197}$Au, respectively. We
fit the ratios of the radiative corrections from nucleus $A$ to carbon
as a linear function of $Z$ to interpolate and extrapolate these
correction factors to the other measured nuclei.  Since the radiative
correction factor changed by only 1\% from Fe to C, we applied the same
factor of 0.741 to all the light nuclei, with a relative uncertainty
to C of 2\%. We used the $^{56}$Fe spectral function for these corrections as the closest published nucleus to $^{54}$Fe.  

The nuclear transparency correction factors, $T_A$, accounted for the probabilities that the struck protons emerged from nuclei without significant rescattering. This correction was determined using the Glauber approximation~\cite{hen14}.  The radiative and transparency correction factors are tabulated in the Supplemental Material.

The systematic uncertainties were determined directly for the single ratios of each nucleus $A$ to C.  Detection efficiencies mostly canceled in these ratios.
The main sources of systematic uncertainties in our analysis were (1) radiative corrections, (2) nuclear transparency corrections, and (3) sensitivity to the choice of analysis cuts (which affected both the measured ratios and the calculated radiative correction factors). The radiative correction uncertainties were dominated by the variation in the ratios due to cut variations.  The transparency uncertainties were estimated from the difference between the Glauber calculated transparency and the transparencies from Ref.~\cite{Duer:2018sjb} and from the size of the correction itself.  

Our cuts were chosen to select SRC events.  However, there is an uncertainty of where exactly to place each cut.  To determine the systematic uncertainty corresponding to this cut uncertainty,   we randomly sampled each cut from a gaussian distribution with an appropriate mean and sigma and calculated the resulting cross-section ratios. We repeated this for many cut choices which resulted in a standard deviation for the cross-section ratios that was taken as the systematic uncertainty.  We applied the same technique to the radiative correction uncertainty.  See Supplemental Material for more details. 

\begin{figure}[htbp]
    \centering
    \includegraphics[width=0.46\textwidth]
    {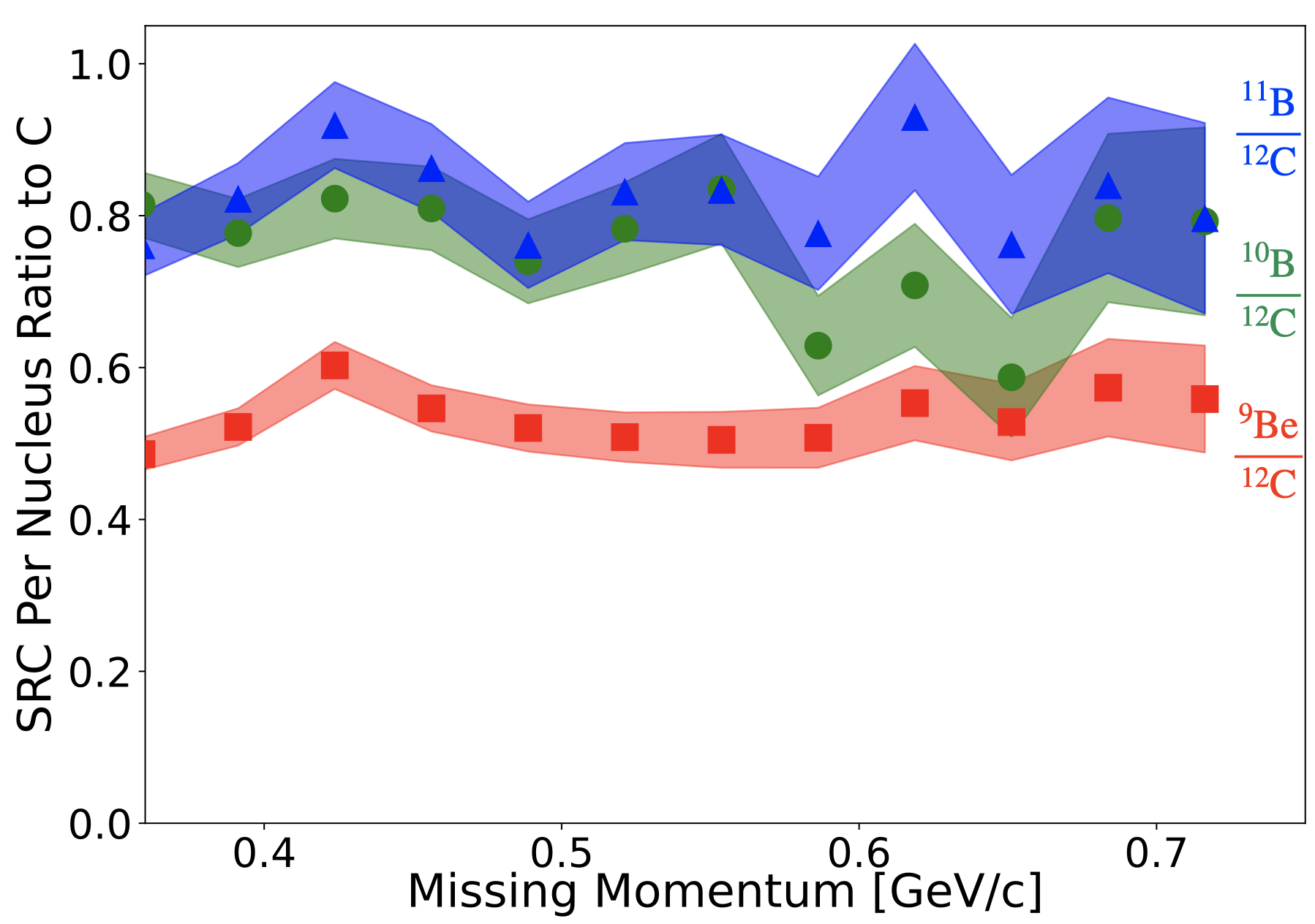}
    \caption[Light Nuclei Missing Momentum for SRC]{The integrated per-nucleus cross-section ratios for $^9$Be, $^{10}$B, and $^{11}$B to C as a function of $p_\mathrm{miss}$.  The red squares show the $^9$Be/C ratio, the green circles show the $^{10}$B/C ratio and the blue triangles show the $^{11}$B/C ratio.  The accompanying bands show the statistical plus systematic uncertainties added in quadrature.} 
    \label{fig:LightPmissRatio}
\end{figure}

For each nucleus, we calculated per-nucleus cross-section ratios to carbon as a function of $p_{\text{miss}}$, see Fig.~\ref{fig:LightPmissRatio}.   The cross-section ratios of the light nuclei to carbon do not depend on $p_\mathrm{miss}$ for $375 \le p_\mathrm{miss}\le 700$ MeV/$c$, consistent with SRC dominance for these kinematics.  The same ratios at lower $p_\mathrm{miss}$ are not constant, reflecting the different shell structures of these nuclei.

\begin{figure}[h!]
    \centering 
\includegraphics[width=0.99\linewidth]{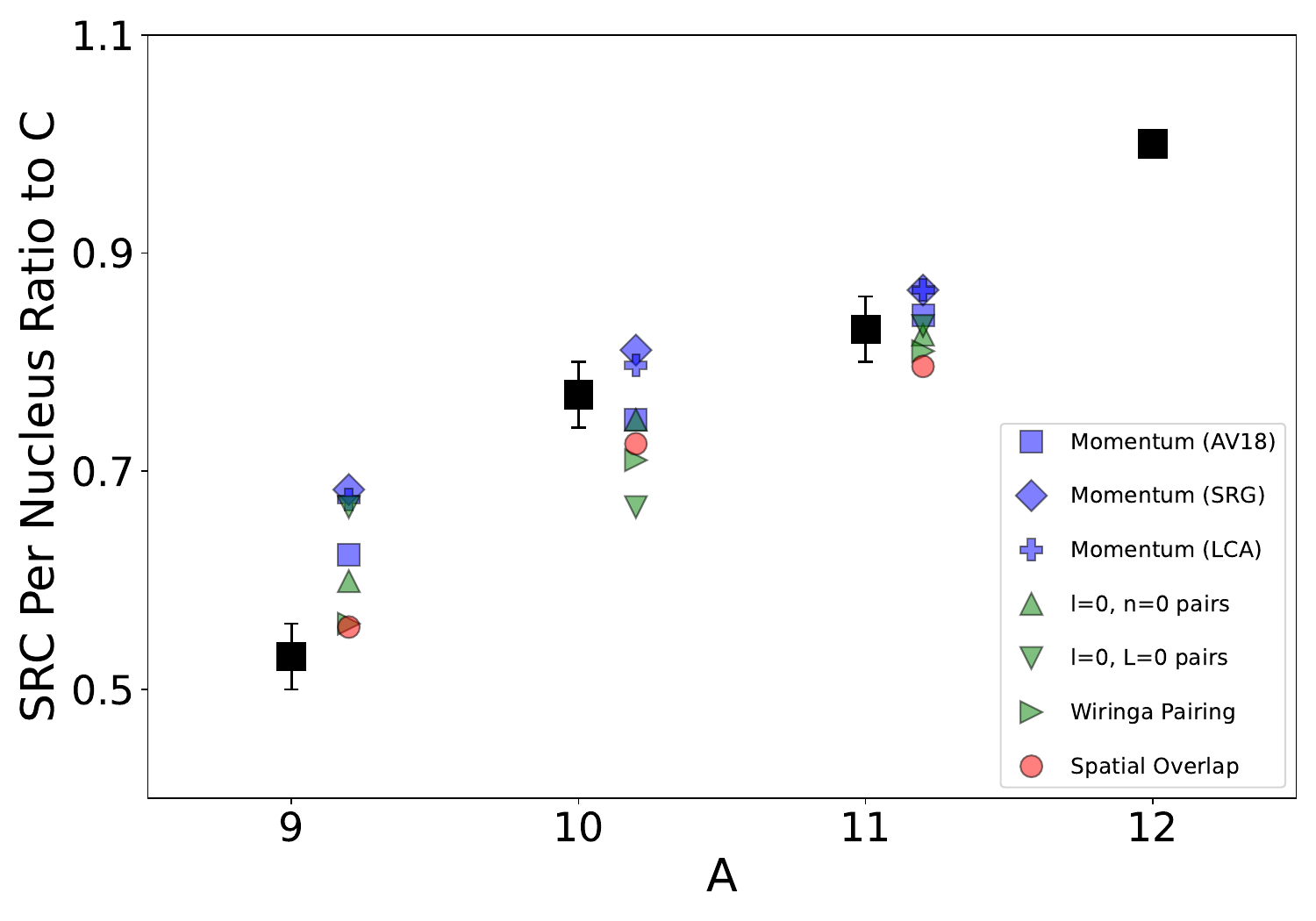}

    \caption{Measured per-nucleus cross section ratio to carbon, $(\sigma_A/Z)/(\sigma_C/6)$, plotted versus $A$ for  $^{9}$Be,
      $^{10}$B, $^{11}$B, and $^{12}$C. The black points show the
      data; the blue squares, diamonds and stars show the quantum montecarlo~\cite{wiringa14}, similarity renormalization group~\cite{Tropiano:2024bmu}, and low-order
      correlation operator approximation~\cite{Ryckebusch:2019oya} momentum
      distribution calculations, respectively; the circles show a
      spatial overlap model~\cite{OverlapModel}; the green upright triangles show the $l=0,n=0$
      pair-counting calculation~\cite{colle15}; and the green inverted triangles
    show the $L=0,l=0$ pair counting model.  The model points are shown slightly offset for clarity.}\label{fig:ratiopernucleus}
\end{figure}

We then integrated the cross-section ratios over $p_\mathrm{miss}$ to obtain
the relative numbers of protons in SRC pairs for nuclei $A$ relative
to C, see Fig.~\ref {fig:ratiopernucleus}.  

We found that the relative
number of SRC protons increased monotonically, almost doubling from
$^9$Be to C.  Comparing $^9$Be to $^{10}$B, we found that adding one
proton increased the number of SRC protons by about 50\%, from 0.53 to
0.77, much greater than the 25\% increase in the number of protons
(from 4 to 5).  We surmise that this added proton could pair with the extra neutron in $^9$Be.  This could also be related to the
large ground state angular momentum ($J=3$) of $^{10}$B, indicating that a valence
$p_{3/2}$ proton and a valence $p_{3/2}$ neutron are maximally
spin-aligned.  Alternatively, it could be due to the fact that $^9$Be can be described in an $\alpha\alpha n$ cluster model~\cite{PhysRevC.54.132} and has an anomalously low average density~\cite{Seely:2009gt,pieper01}.  

\begin{figure}[htbp]
    \centering \includegraphics[width=1.0\linewidth]{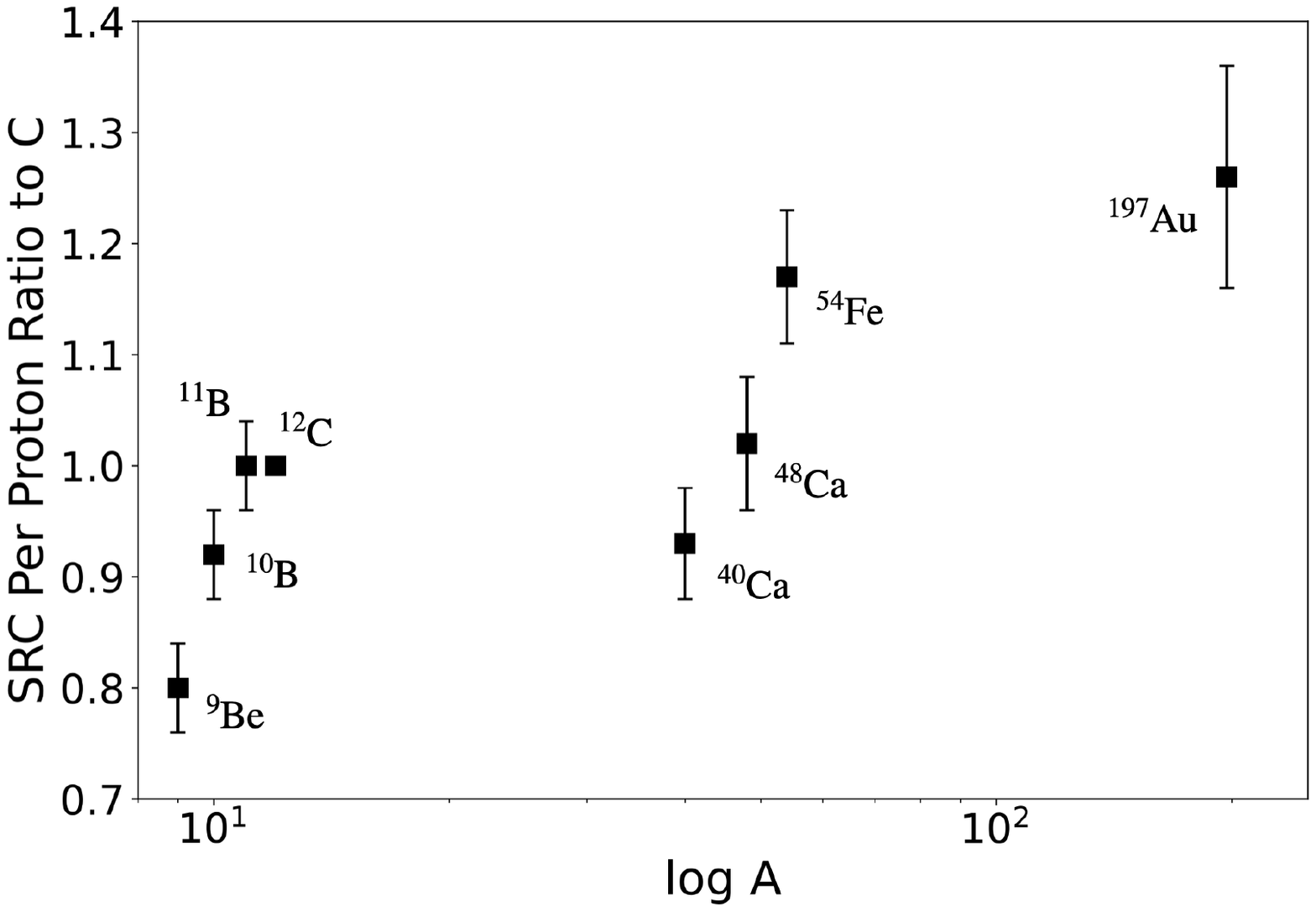}
    \caption{Measured per-proton cross section ratio to carbon, $(\sigma_A/Z)/(\sigma_C/6)$, plotted versus  $A$ for  $^{9}$Be, $^{10}$B, $^{11}$B, $^{12}$C, $^{40}$Ca,  $^{48}$Ca,  $^{54}$Fe, and  $^{197}$Au. The cross-section ratios shown in this plot are per-proton, for ease of plotting.}\label{fig:ratiovsA}
\end{figure}

Similarly, comparing $^{10}$B to $^{11}$B, we found that adding a neutron increased the relative number of SRC protons from 0.77 to 0.83.  While only a two-sigma effect, this shows that increasing the number of neutrons can increase the pairing of protons~\cite{duer18}.  Finally, comparing $^{11}$B to $^{12}$C, we found that adding a proton increased the number of SRC protons by 20\%, identical to the 20\% increase in $Z$.

The SRC pairing increase from $^9$Be ($J=3/2$) to $^{10}$B ($J=3$) is much larger than that from $^{10}$B to $^{12}$C ($J=0$), implying that the total angular momentum is less important than the cluster structure of $^9$Be.

When we compared the relative fraction of SRC protons for all measured
nuclei from $^9$Be to $^{197}$Au, we found that it did not depend on $N/Z$
and that the $A$-dependent slope was much steeper within the light
nuclei ($9\le A\le 12$) and within the medium nuclei ($40\le A\le 54$)
than overall ($9\le A\le 197$), see Fig.~\ref{fig:ratiovsA}.  This provides more evidence that the SRC-pair probability is strongly modified by the details of low-momentum nuclear structure.

\begin{table}[htbp!]
 \begin{center}
 \setlength{\tabcolsep}{2.pt}
 \small
 \begin{tabular}{| l | c | l | l | l | l | l | l|l |} \hline
     
     Target
      &  Data  &  AV18  &  SRG  &  LCA  &  Spa  & WP
      & \begin{tabular}[c]{@{}l@{}}$l=0$\\$n=0$\end{tabular}
      & \begin{tabular}[c]{@{}l@{}}$l=0$ \\$L=0$\end{tabular} \\ \hline    
    $^{9}$Be & 0.53 $\pm$ 0.03 & 0.62 & 0.68 & 0.67 & 0.56  &0.56 & 0.60 & 0.67 \\ \hline
    $^{10}$B & 0.77 $\pm$ 0.03 & 0.75 & 0.81 & 0.80 & 0.71  &0.71 & 0.75 & 0.67 \\ \hline
    $^{11}$B & 0.83 $\pm$ 0.03 & 0.84 & 0.87 & 0.87 & 0.81  &0.81 & 0.82 & 0.83 \\ \hline
    $^{12}$C & 1               & 1     & 1    & 1    & 1     & 1   &  1   & 1\\ \hline
    \end{tabular}
    \end{center}
    \caption{Per-nucleus cross-section ratios for different nuclei to
      C, $\sigma_A/\sigma_C$, for both data and models.  AV18, SRG and LCA refer to the the quantum montecarlo~\cite{Lonardoni:2017egu}, similarity renormalization group~\cite{Tropiano:2024bmu}, and low-order
      correlation operator approximation~\cite{Ryckebusch:2019oya} momentum
      distribution calculations, respectively.  ``Spa'' refers to the spatial overlap calculation~\cite{OverlapModel}. ``WP'' refers to the Wiringa spin-isospin pairing model~\cite{Wiringa:2006ih}.  $l=0,n=0$~\cite{colle15}and $l=0,L=0$ refer to the respective pair counting models.} 
    \label{data-model}
\end{table}

We compared the integrated cross-section ratio  to several
theoretical models, see Fig.~\ref{fig:ratiopernucleus}.  The proton
momentum distribution model results were determined by integrating the calculated
momentum distributions over $375\le p_{\text{miss}}\le 700$ MeV/$c$
and forming the ratio to $^{12}$C. However, since we do not have calculated spectral
functions (i.e., two-dimensional distributions in missing energy and
missing momentum), this comparison is not exact.


The AV18 momentum distributions were calculated from the nuclear wave
function obtained by solving the nuclear many-body problem using the
AV18 nucleon-nucleon interaction~\cite{wiringa14} for
$^{9}$Be, $^{10}$B, $^{11}$B, $^{12}$C and $^{40}$Ca. Tropiano {\it et
  al.} calculated the momentum distributions of all measured nuclei in
this experiment using similarity renormalization group (SRG) evolved
operators and empirically fit single-particle
orbitals~\cite{Tropiano:2024bmu}. Ryckebusch \textit{et
  al.}~\cite{Ryckebusch:2019oya} used the low-order correlation
operator approximation (LCA) to compute the SRC contribution to the
single-nucleon momentum distribution.
All three models,   AV18, SRG, and LCA, described the data very well  for $^{10}$B and $^{11}$B, but overestimated $^9$Be by 15 to 25\%.

The spatial overlap model~\cite{OverlapModel} used Woods-Saxon
shell-model wave functions with Cohen-Kurath occupancies to calculate the relative overlap
probabilities for $pn$ and $pp$ pairs in each nucleus.  It included a suppression factor for $pp$ relative to $np$ pairs due to tensor force
dominance.  It described  the relative number of SRC protons in all nuclei remarkably well. 

We also compared our results with three quantum pairing models.   Vanhalst, Ryckebush and Cosyn~\cite{Vanhalst:2011es,vanhalst12,colle15} calculated the  number of $pp$ and $pn$ SRC pairs in nuclei for nucleons in a relative nodeless ($n=0$)  $S$ state (i.e., with zero relative angular momentum, $l=0$). The number of high-$p_\mathrm{miss}$ protons is then proportional to the number of $pn$ pairs plus twice the number of $pp$ pairs.
This quantum selection rule described the data 
well  for $^{10}$B and $^{11}$B, but overestimated $^9$Be by 13\%.

Wiringa~\cite{Wiringa:2006ih} calculated the number of spin-isospin $NN$  pairs in light nuclei to estimate binding energies and excited states.  He found that $sp$ pairs did not contribute to the binding energy.  We used the relative number of spin-1, isospin-0 $np$ pairs in light nuclei as a proxy for the number of SRC pairs.  This agreed remarkably well with the measured data.

We also calculated the relative number of $np$ SRC pairs by requiring
both relative and total pair angular momentum to be zero, $l=0$ and
$L=0$ \cite{Lane1964}.   The $L=0$ requirement eliminates pairing
between shells with different orbital angular momenta. This model also
assumed that the $NN$ correlation function was only significant for
small values of the relative position, equivalent to a zero-range
approximation. It predicts that the ratio of $np$ SRC-pairs in
$^{10}$B/$^{11}$B is 0.8, since the ground state spin of $^{10}$B is
3, indicating that one of its five protons cannot couple to $L=0$.
This model predicts the same number of  SRC protons in  $^9$Be
and  $^{10}$B, which is contradicted by the 50\% increase seen in the data.

These results showcase the tremendous progress in precise calculations to predict the high-momentum structure of nuclei and the importance of using  ratios to isolate interesting aspects of the nucleus.  The momentum-distribution calculations agreed with the measured ratios of the boron nuclei to carbon to within the experimental uncertainties.  They also agreed within 15-20\% to the Be/C ratio.

Surprisingly, two simpler models, the spatial overlap model and the Wiringa pair-counting model, both described the data better than more sophisticated models.

Our measurements of SRC-pair relative probabilities are qualitatively similar to the EMC effect measurements of~\cite{HallC:2022utd}.  Both measurements appear to see a larger change (either in the EMC slope or the SRC-pair probability) from $^9$Be to $^{10}$B, than from $^{10}$B to $^{11}$B or from $^{11}$B to $^{12}$C, although the EMC measurements have larger relative uncertainties.  This similarity is consistent with the strong correlation between the strength of the EMC effect and the relative fraction of SRC pairs in nuclei~\cite{PhysRevLett.106.052301,RevModPhys.89.045002}.  

In conclusion, we investigated details of  SRC pairing in light nuclei.  Adding one proton to $^9$Be increased the relative number of SRC protons far more than adding  one neutron to $^{10}$B or one  proton to $^{11}$B, perhaps due to the unusual cluster structure of $^9$Be or the $J=3$ ground state of $^{10}$B.   In addition, the fraction of high-initial-momentum protons increased much more rapidly from $A=9$ to 12 and from $A=40$ to 54 than from $A=9$ to 197. These results complement the  shell effects  seen in $^{40}$Ca, $^{48}$Ca, and
$^{54}$Fe where \textit{intra}-shell pairing was much stronger than \textit{inter}-shell pairing~\cite{CaFeNature}.  
Together, these shows the importance of low-momentum structure for high-momentum SRC pairing.

\begin{acknowledgments}
We acknowledge the efforts of the staff of the Accelerator and Physics Divisions at Jefferson Lab that made this experiment possible. This work was produced in part by SURATech, LLC under Contract No. 892431-26-C-SC000213 with the U.S. Department of Energy, and support from the National Science Foundation MPS-Ascend Postdoctoral Research Fellowship Grant No. 2137604 (CY), the U.S. Department of Energy, D.O.E. grants DE-SC0022007 (HSV), DE-SC0020240 (OH, JK and AD), DE-FG02-96ER40960 (NS, CF, CAG and LBW), DE-SC0016583 (AS, PS), DE-SC0013615 (CM), DE-AC02-05CH11231 (JK via the Lawrence Berkeley National Laboratory), the Jefferson Lab Nathan Isgur Fellowship, UTK-JLab bridge position (DN),  the United States-Israel Binational Science Foundation (BSF) grant 2024036 (EP), the Israeli Science Foundation grants 917/20, 371/23 (EP), and 830/24 (IK), and the PAZY Foundation grant 520/23 (EP). 
\end{acknowledgments}

\medskip
The raw data from this experiment are archived in the Jefferson Lab’s
mass storage silo under the Hall C dataset. Access to these data can
be facilitated by contacting either the corresponding author or the
Jefferson Lab computing centre at helpdesk@jlab.org. Source data
are provided with this paper.

\bibliography{references.bib}


\end{document}


\title{Supplementary Materials: Short-range correlated pair formation and nuclear shell structure}

\pacs{}
\maketitle

\section{Kinematic Coverage and event selection cuts}
\begin{figure}[ht!]
    \centering
    \includegraphics[width=0.8\linewidth]{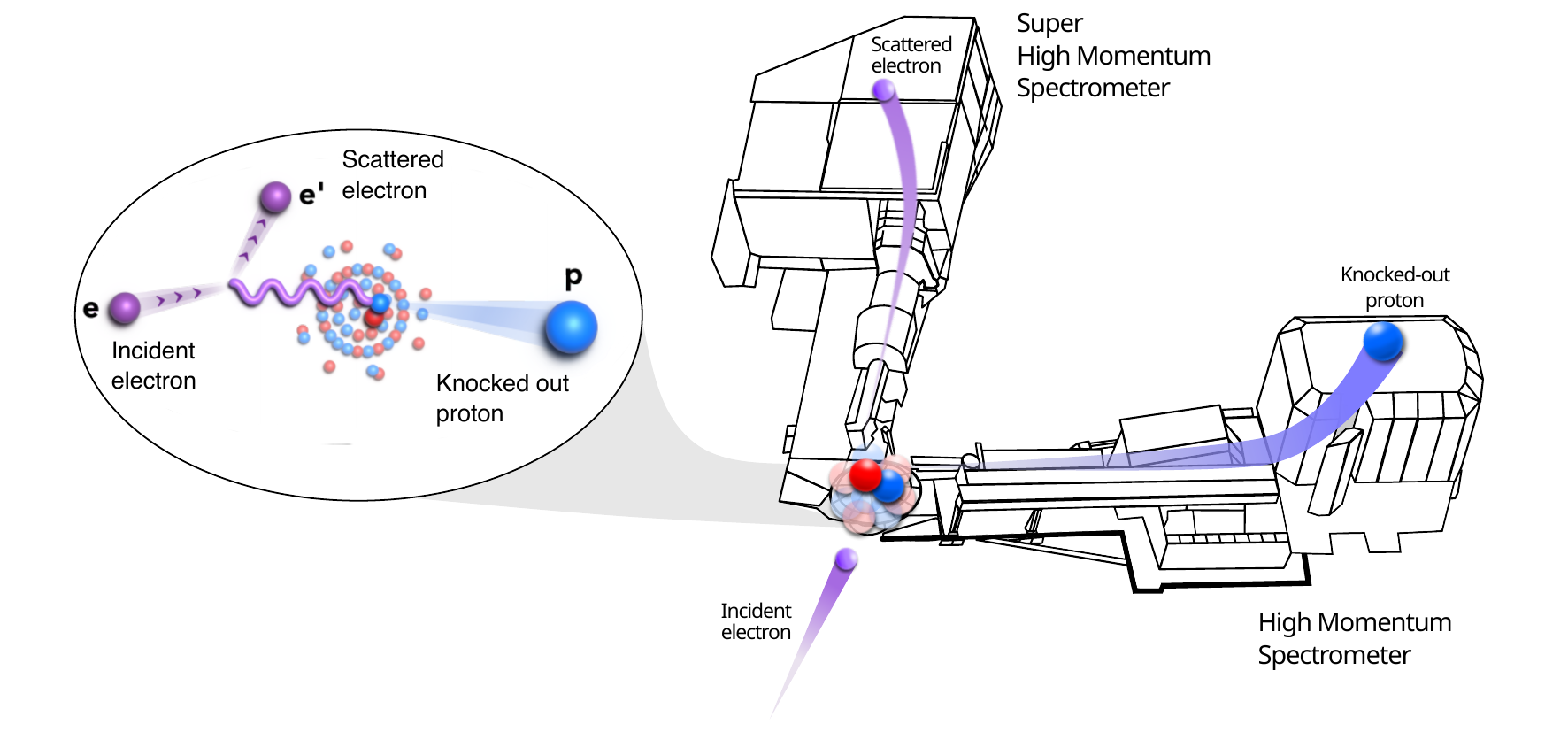}
    \caption{A schematic of the Hall C Super High Momentum Spectrometer and High Momentum Spectrometers.  The scattered electrons were detected in the SHMS, and the knocked-out protons were detected in the HMS}
    \label{spectrometer}
\end{figure}

We measured scattered electrons and knocked-out protons in the Super High Momentum Spectrometer (SHMS) and the High Momentum Spectrometer (HMS), respectively (see Fig.~\ref{spectrometer}). Table~\ref{central-kin} shows the central angle and momentum settings of both spectrometers.  

\begin{figure}[!h]
   \includegraphics[width=0.45\textwidth]{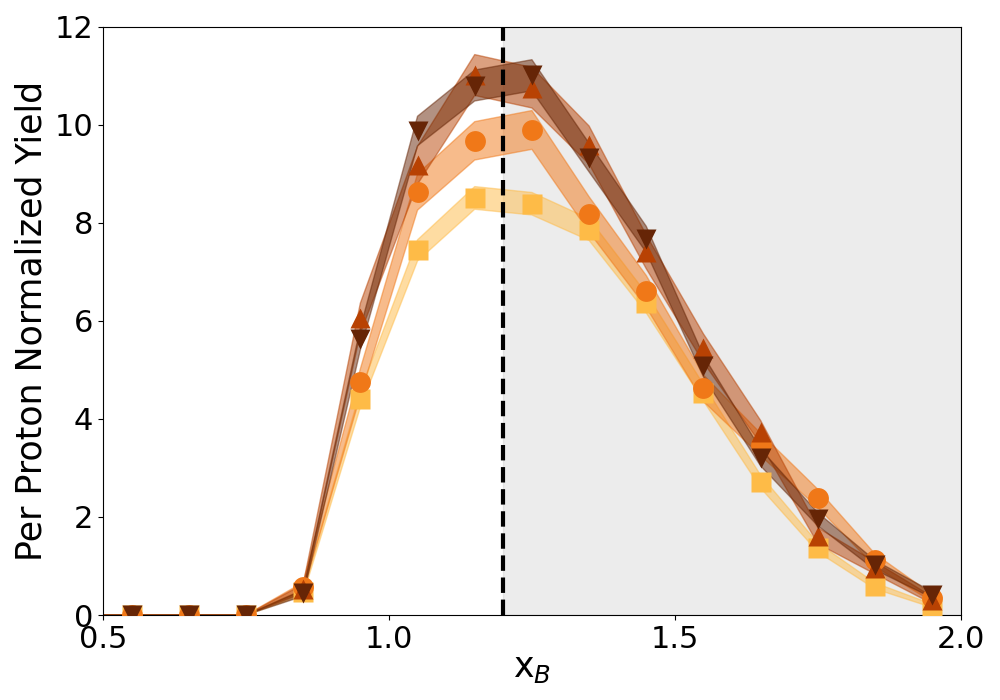}
   \includegraphics[width=0.45\textwidth]{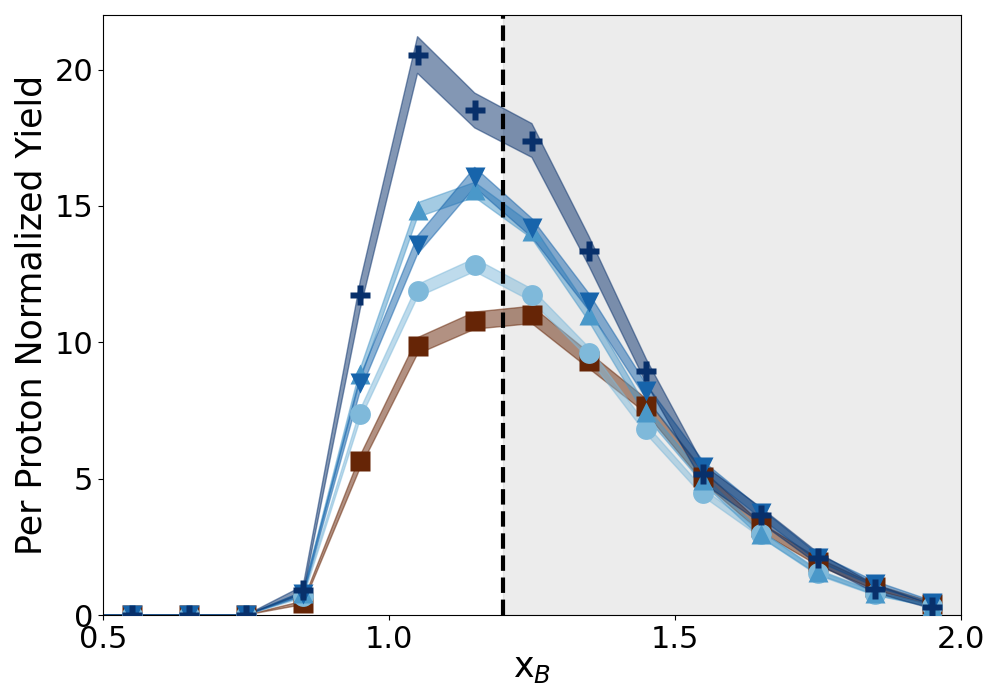}
   \caption{Normalized yield plotted versus $x_{bj}$. The left plot is for  $^{9}$Be, $^{10}$B, $^{11}$B and $^{12}$C. The right plot is for $^{12}$C, $^{40}$Ca, $^{48}$Ca, $^{54}$Fe and $^{197}$Au. The colors get darker with increasing $A$ for both plots.  Carbon (dark brown) is shown in both plots for comparison. The shaded region indicates the event selection cut on $x_{bj} > 1.2$.}
   \label{kin-x}
\end{figure}

\begin{figure}[!h]
   \includegraphics[width=0.45\textwidth]{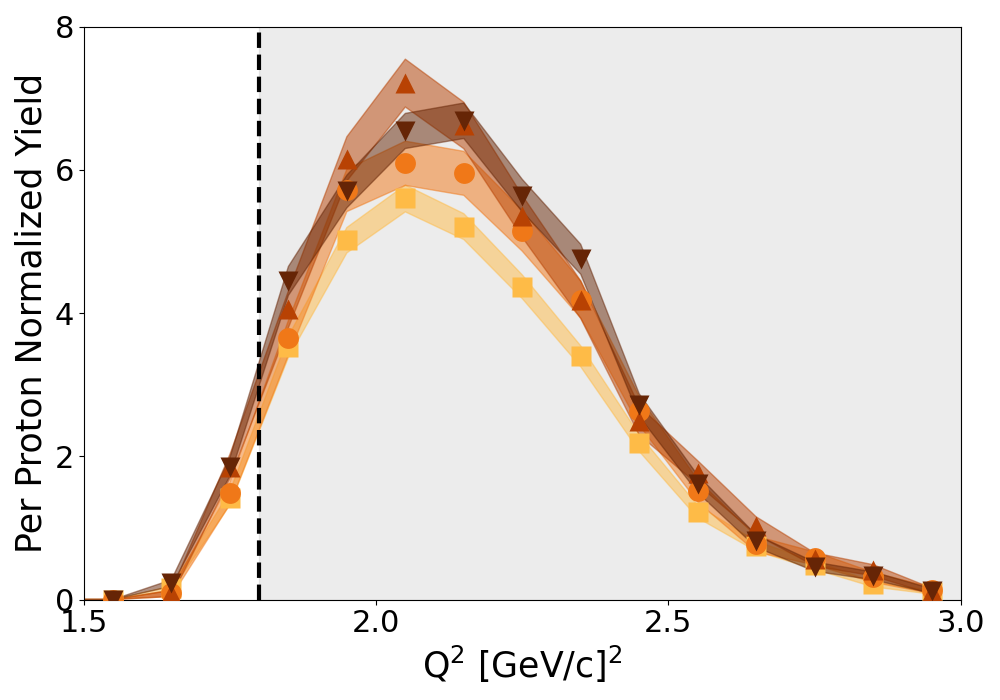}
   \includegraphics[width=0.45\textwidth]{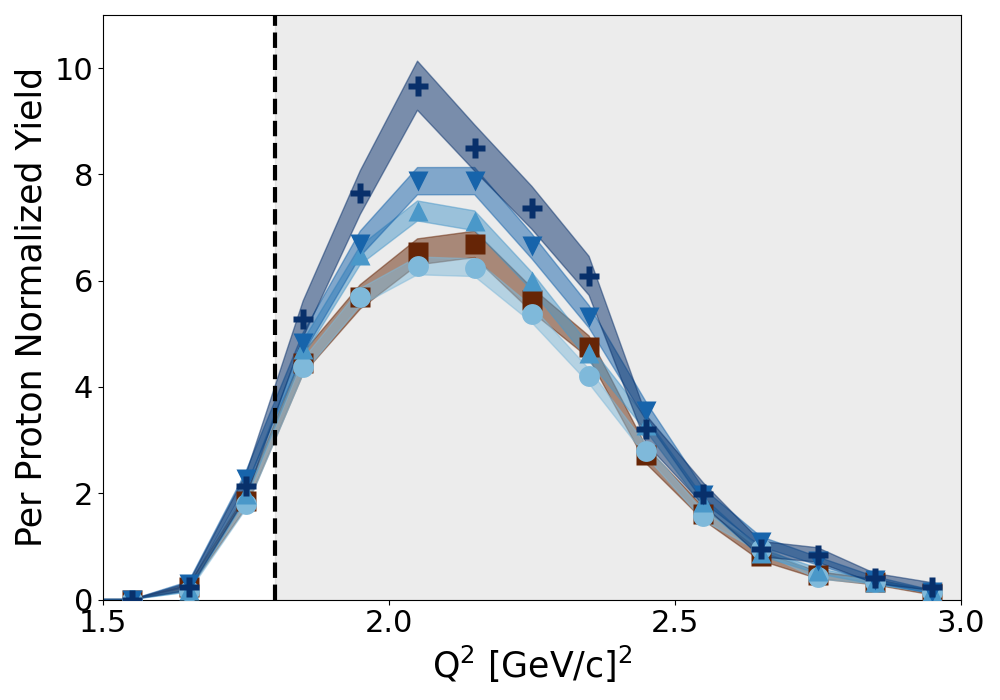}
   \caption{Normalized yield plotted versus $Q^2$. The left plot is for  $^{9}$Be, $^{10}$B, $^{11}$B and $^{12}$C. The right plot is for $^{12}$C, $^{40}$Ca, $^{48}$Ca, $^{54}$Fe and $^{197}$Au. The colors get darker with increasing $A$ for both plots.  Carbon (dark brown) is shown in both plots for comparison. The shaded region indicates the event selection cut on $Q^2 > 1.8$.}
   \label{kin-q2}
\end{figure}

\begin{figure}[!h]
   \includegraphics[width=0.45\textwidth]{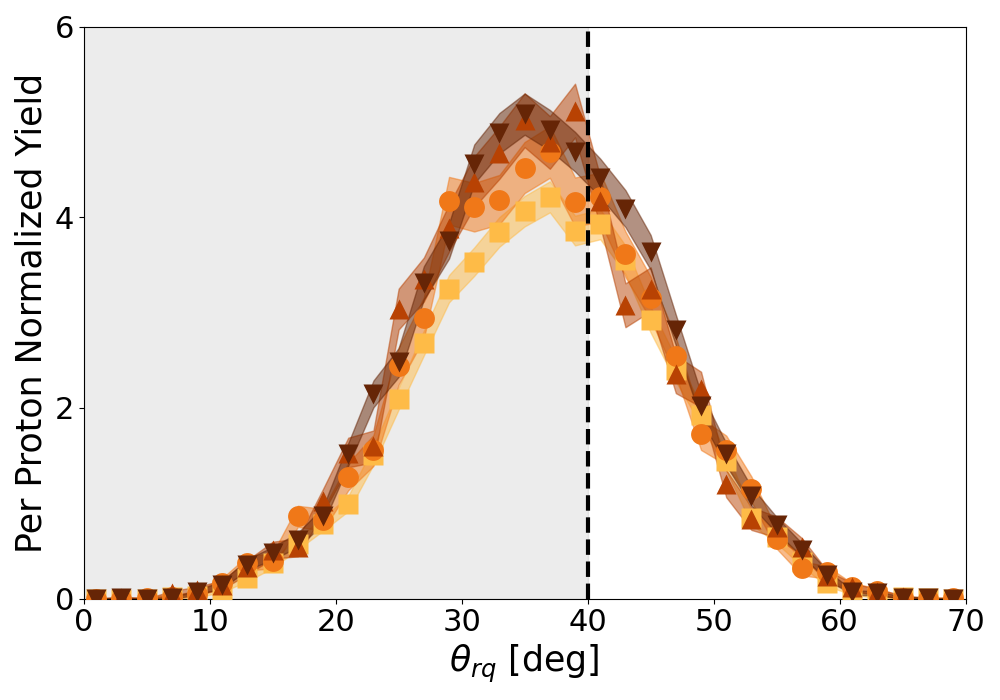}
   \includegraphics[width=0.45\textwidth]{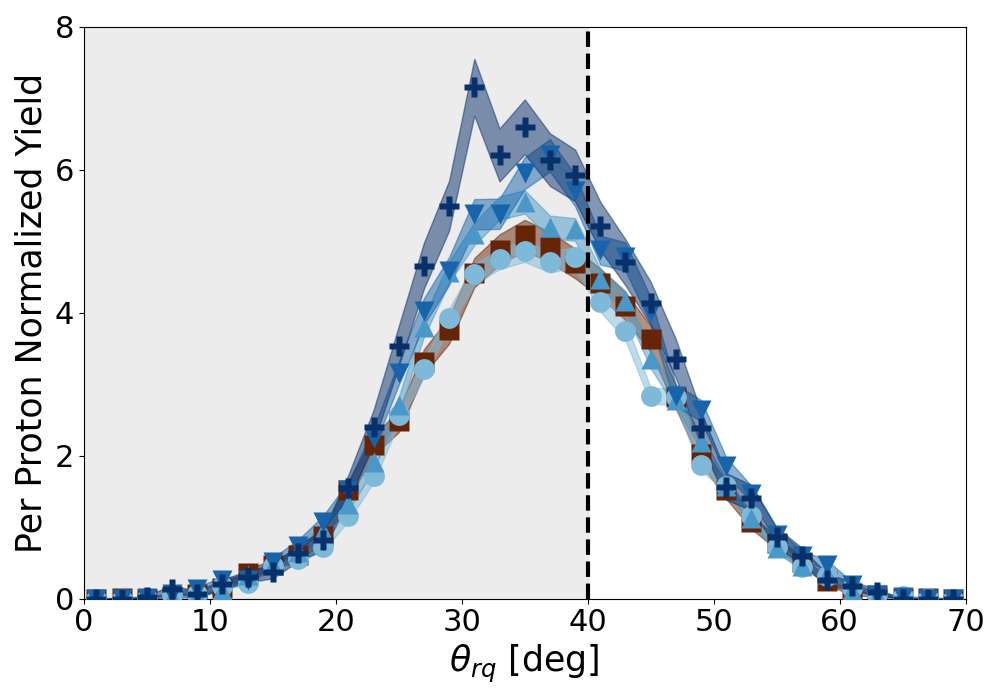}
   \caption{Normalized yield plotted versus $\theta_{rq}$. The left plot is for  $^{9}$Be, $^{10}$B, $^{11}$B and $^{12}$C. The right plot is for $^{12}$C, $^{40}$Ca, $^{48}$Ca, $^{54}$Fe and $^{197}$Au. The colors get darker with increasing $A$ for both plots. Carbon (dark brown) is shown in both plots for comparison. The shaded region indicates the event selection cut on $\theta_{rq} < 40^{\circ}$.}
   \label{kin-thetarq}
\end{figure}

\begin{figure}[!h]
   \includegraphics[width=0.45\textwidth]{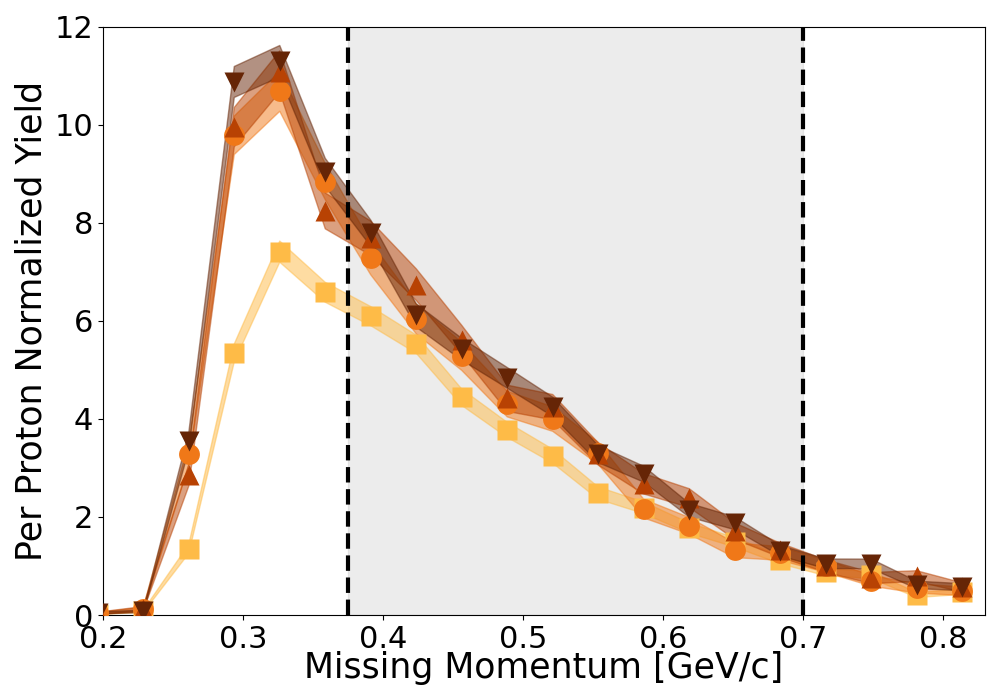}
   \includegraphics[width=0.45\textwidth]{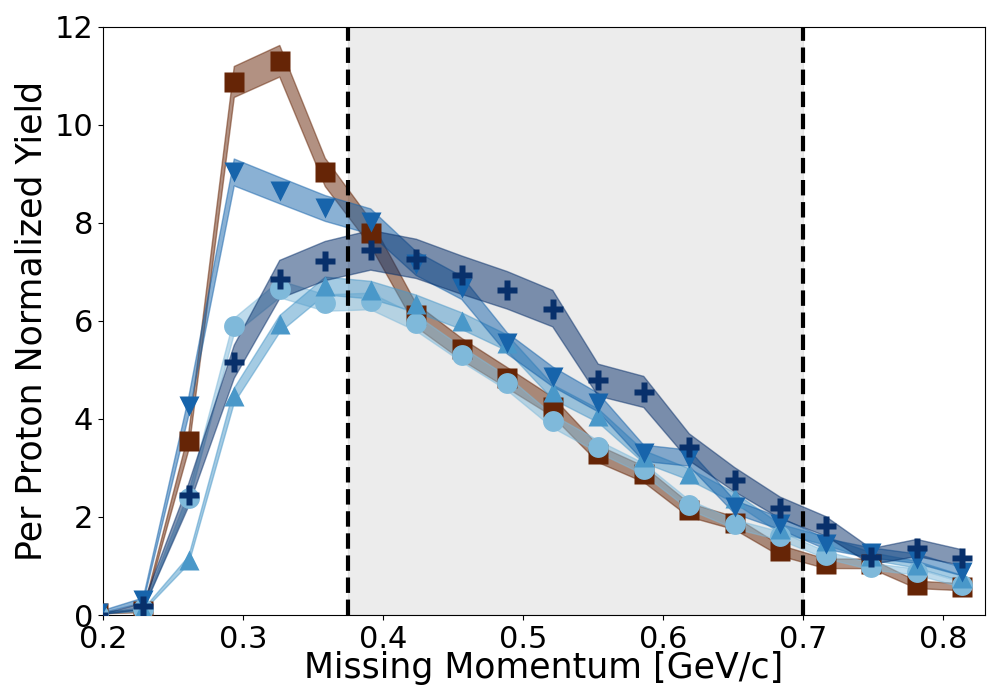}
   \caption{Normalized yield plotted versus $Q^2$. The left plot is for  $^{9}$Be, $^{10}$B, $^{11}$B and $^{12}$C. The right plot is for $^{12}$C, $^{40}$Ca, $^{48}$Ca, $^{54}$Fe and $^{197}$Au. The colors get darker with increasing $A$ for both plots.  Carbon (dark brown) is shown in both plots for comparison. The shaded region indicates the event selection cut on $375~\text{MeV}/c < P_{miss} < 700~\text{MeV}/c$.}
   \label{kin-pmiss}
\end{figure}

\begin{figure}[!h]
   \includegraphics[width=0.45\textwidth]{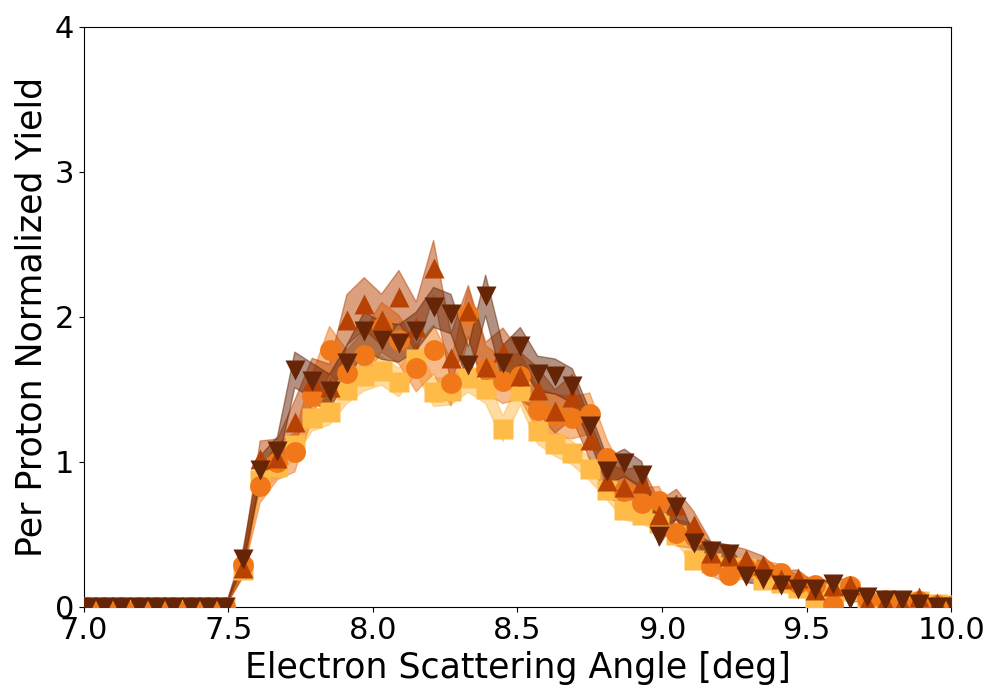}
   \includegraphics[width=0.45\textwidth]{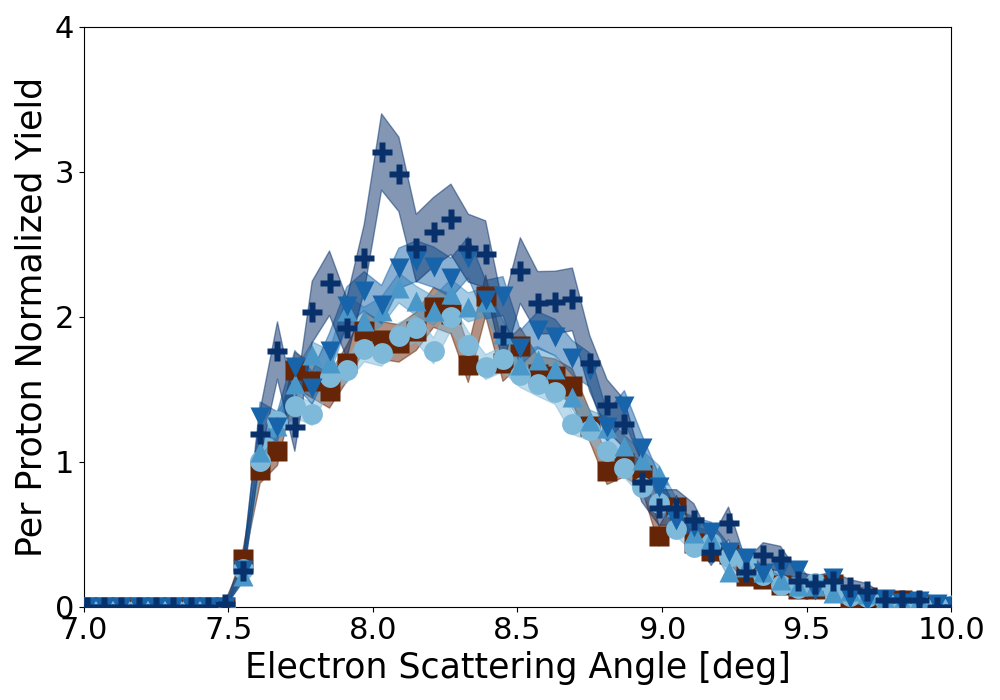}
   \caption{Normalized yield plotted versus electron scattering angle $\theta_e$. The left plot is for  $^{9}$Be, $^{10}$B, $^{11}$B and $^{12}$C. The right plot is for $^{12}$C, $^{40}$Ca, $^{48}$Ca, $^{54}$Fe and $^{197}$Au. The colors get darker with increasing $A$ for both plots.  Carbon (dark brown) is shown in both plots for comparison. }
   \label{kin-the}
\end{figure}

\begin{figure}[!h]
   \includegraphics[width=0.45\textwidth]{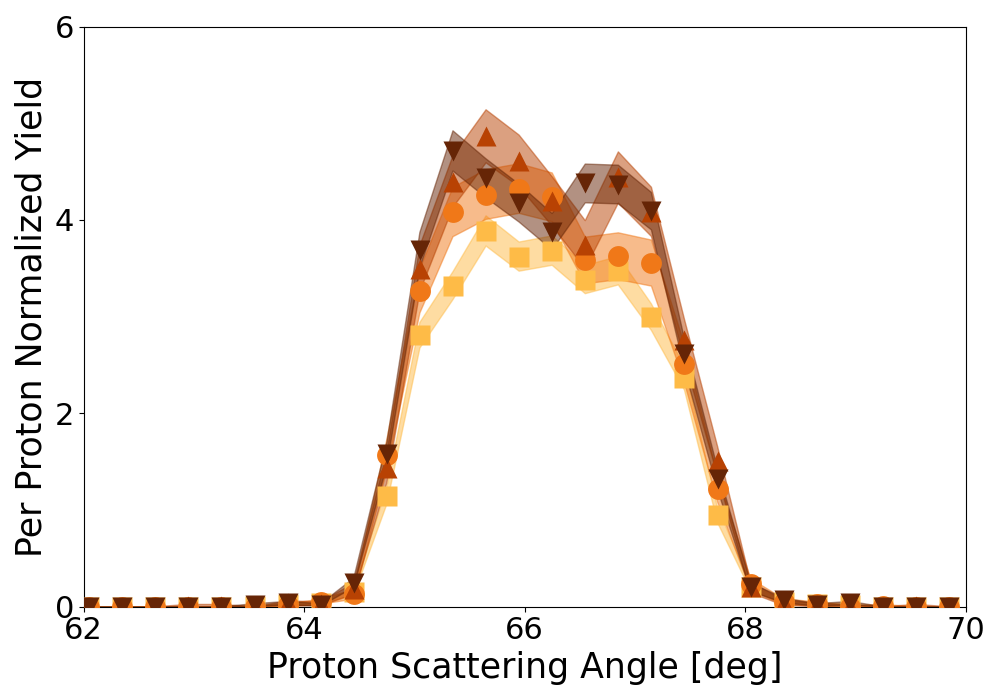}
   \includegraphics[width=0.45\textwidth]{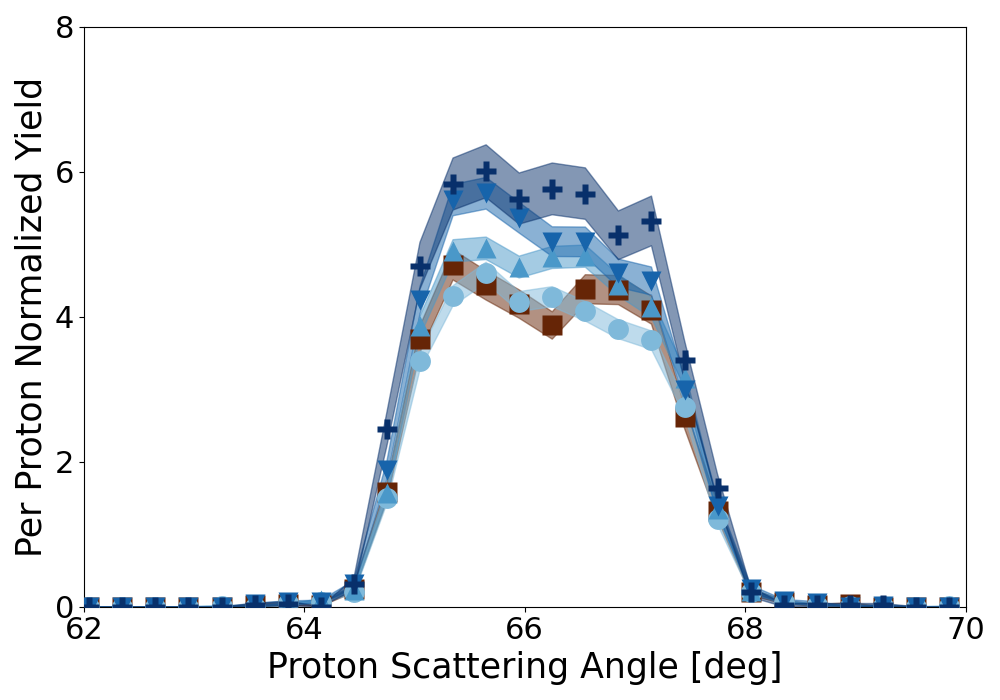}
   \caption{Normalized yield plotted versus proton scattering angle $\theta_p$. The left plot is for  $^{9}$Be, $^{10}$B, $^{11}$B and $^{12}$C. The right plot is for $^{12}$C, $^{40}$Ca, $^{48}$Ca, $^{54}$Fe and $^{197}$Au. The colors get darker with increasing $A$ for both plots.  Carbon (dark brown) is shown in both plots for comparison. }
   \label{kin-thp}
\end{figure}

\begin{figure}[!h]
   \includegraphics[width=0.45\textwidth]{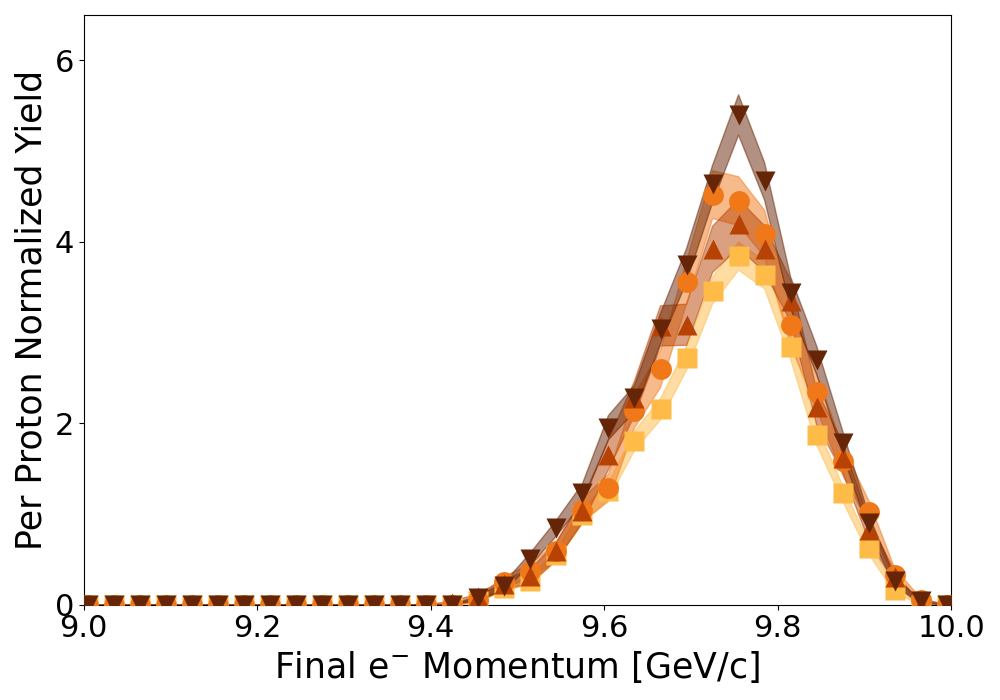}
   \includegraphics[width=0.45\textwidth]{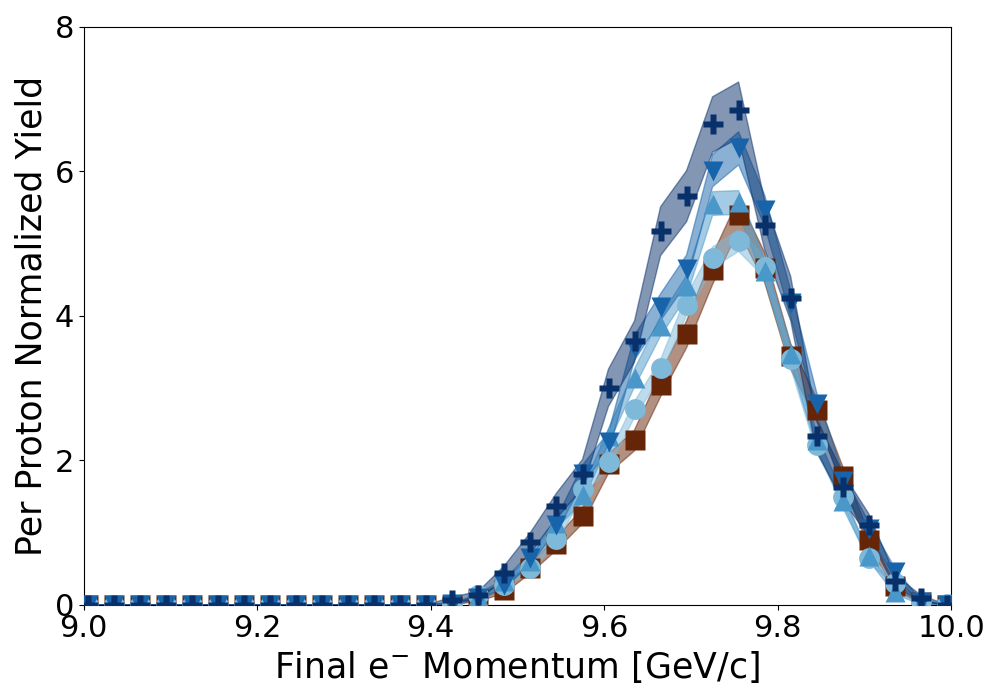}
   \caption{Normalized yield plotted versus electron scattering momentum $p_e$. The left plot is for  $^{9}$Be, $^{10}$B, $^{11}$B and $^{12}$C. The right plot is for $^{12}$C, $^{40}$Ca, $^{48}$Ca, $^{54}$Fe and $^{197}$Au. The colors get darker with increasing $A$ for both plots.  Carbon (dark brown) is shown in both plots for comparison. }
   \label{kin-pe}
\end{figure}

\begin{figure}[!h]
   \includegraphics[width=0.45\textwidth]{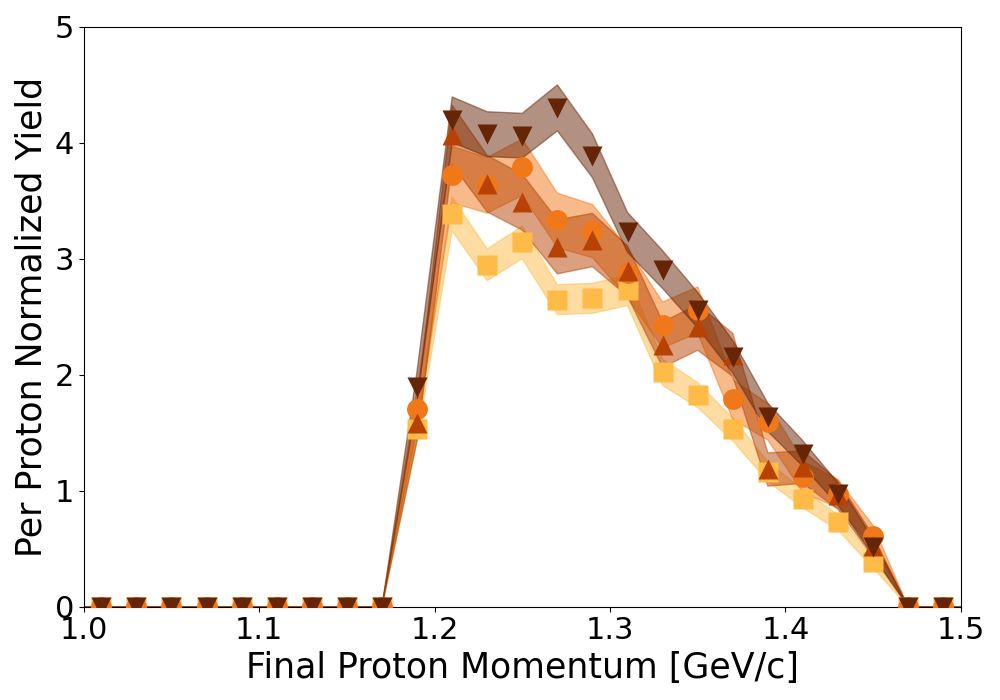}
   \includegraphics[width=0.45\textwidth]{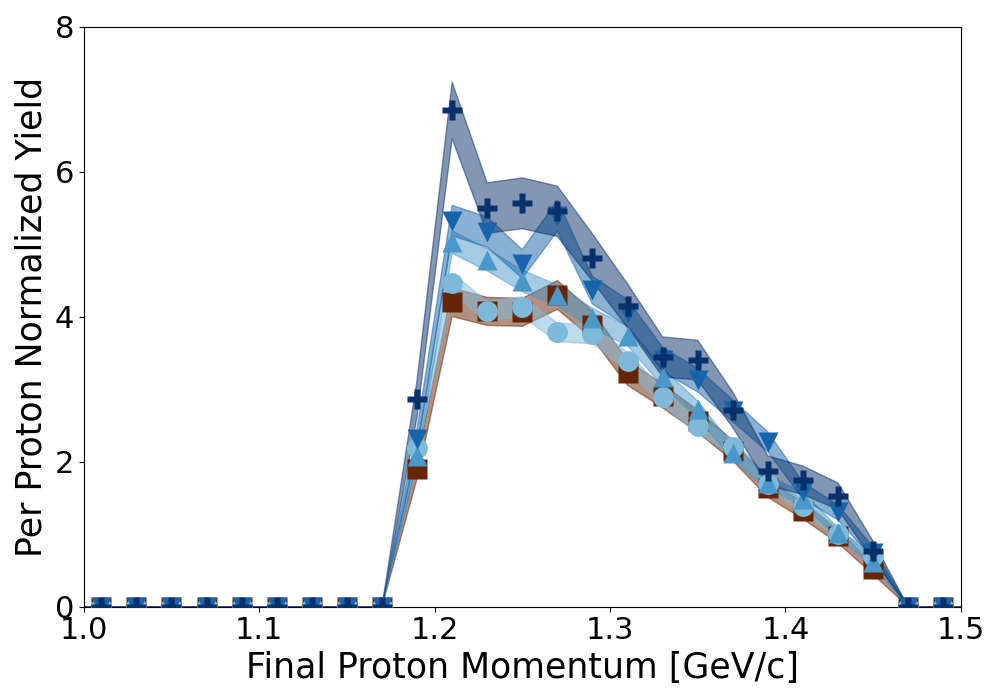}
   \caption{Normalized yield plotted versus proton scattering momentum $p_p$. The left plot is for  $^{9}$Be, $^{10}$B, $^{11}$B and $^{12}$C. The right plot is for $^{12}$C, $^{40}$Ca, $^{48}$Ca, $^{54}$Fe and $^{197}$Au. The colors get darker with increasing $A$ for both plots.  Carbon (dark brown) is shown in both plots for comparison. }
   \label{kin-pp}
\end{figure}

The kinematic coverage for $x_b$, $Q^2$, $\theta_{rq}$, $P_{miss}$ are shown in Figs.~\ref{kin-x} to ~\ref{kin-pmiss}. The shaded regions indicate the event selection cuts.  The resulting distributions of the electron and proton momenta and angles are shown in Figs.~\ref{kin-the} to \ref{kin-pp}.

\begin{table}[ht!]
    \centering
    \begin{tabular}{|c|c|c|c|}
        \hline
        $\theta_e$  & $P_e$ [GeV] & $\theta_p$  & $P_p$ [GeV] \\ \hline\hline
         8.3$^{\circ}$ & 8.55  & 66.4$^{\circ}$ & 1.325  \\ \hline
    \end{tabular}
    \caption{Central kinematic settings of the SHMS and HMS spectrometers.  The SHMS momentum acceptance ranged between $-10\%$ and $+22\%$ of the central momentum.  The  HMS momentum acceptance ranged between $-10\%$ and $+10\%$ of the central momentum.}
    \label{central-kin} 
\end{table}

\section{Target information}
The target information  is summarized in Table~\ref{target}.  The boron targets were made of B$_4$C.  The C contribution was measured using the C target and subtracted from the B$_4$C data to get the B results.  The ``$^{48}$Ca'' target contained 10\% $^{40}$Ca.  This contribution was measured using the $^{40}$Ca target and subtracted from the  ``$^{48}$Ca'' target data to get the $^{48}$Ca results.

\begin{table}[h!]
  \begin{center}
    \begin{tabular}{|l|c|r|}\hline
      Target & Thickness (g/cm$^2$) & purity\\ \hline
      $^9$Be & 0.978 & 100\% \\\hline
      $^{10}$B   & 0.5722 & 96.6\% \\\hline
      $^{11}$B   & 0.6344 & 99.8\% \\\hline
      $^{12}$C  & 0.5244 & 99.9\% \\\hline
      $^{40}$Ca & 0.800 & 100\% \\\hline
      $^{48}$Ca & 1.050 & 90\% \\\hline
      $^{54}$Fe & 0.4152 & 98\% \\\hline
      $^{197}$Au & 0.388 & 100\% \\\hline
    \end{tabular} 
    \label{target}
    \caption{Target areal densities and purities.  The beryllium, carbon, and gold targets were of natural abundances. } 
    \end{center}
\end{table}
\subsection*{Oil contamination for Calcium targets}
Both Ca targets were stored in  light mineral oil (typically (CH$_2$)$_n$) to prevent oxidation. During the experiment it was discovered that the $^{40}$Ca and $^{48}$Ca targets had a layer of mineral oil which slowly evaporated with time and beam exposure. We measured the target oil contamination in two ways. First, we measured the H$(e,e'p)$ peak at $E_{miss}=0$ and $p_{miss}=0$ for  the Ca$(e,e'p)$ calibration runs before and after the data taking to directly measure the hydrogen contamination. Second, we used the rate of SHMS single-arm electron triggers per incident electron for Ca$(e,e')$ to measure the total target thickness (Ca plus oil contamination) in each run. The $^{48}$Ca oil contamination decreased exponentially from about 3\% to about 0.5\% during the data taking as the oil evaporated in the vacuum of the target chamber. The $^{40}$Ca oil contamination was constant at about 0.5\%. These total oil contamination values were consistent with the measured H contamination for (CH$_2$)$_n$ mineral oil. We subtracted the oil contamination run-by-run using our measured C$(e,e'p)$ data, scaled to the estimated carbon content of the oil.

\begin{figure}[ht!]
    \centering
    \includegraphics[width=0.75\linewidth]{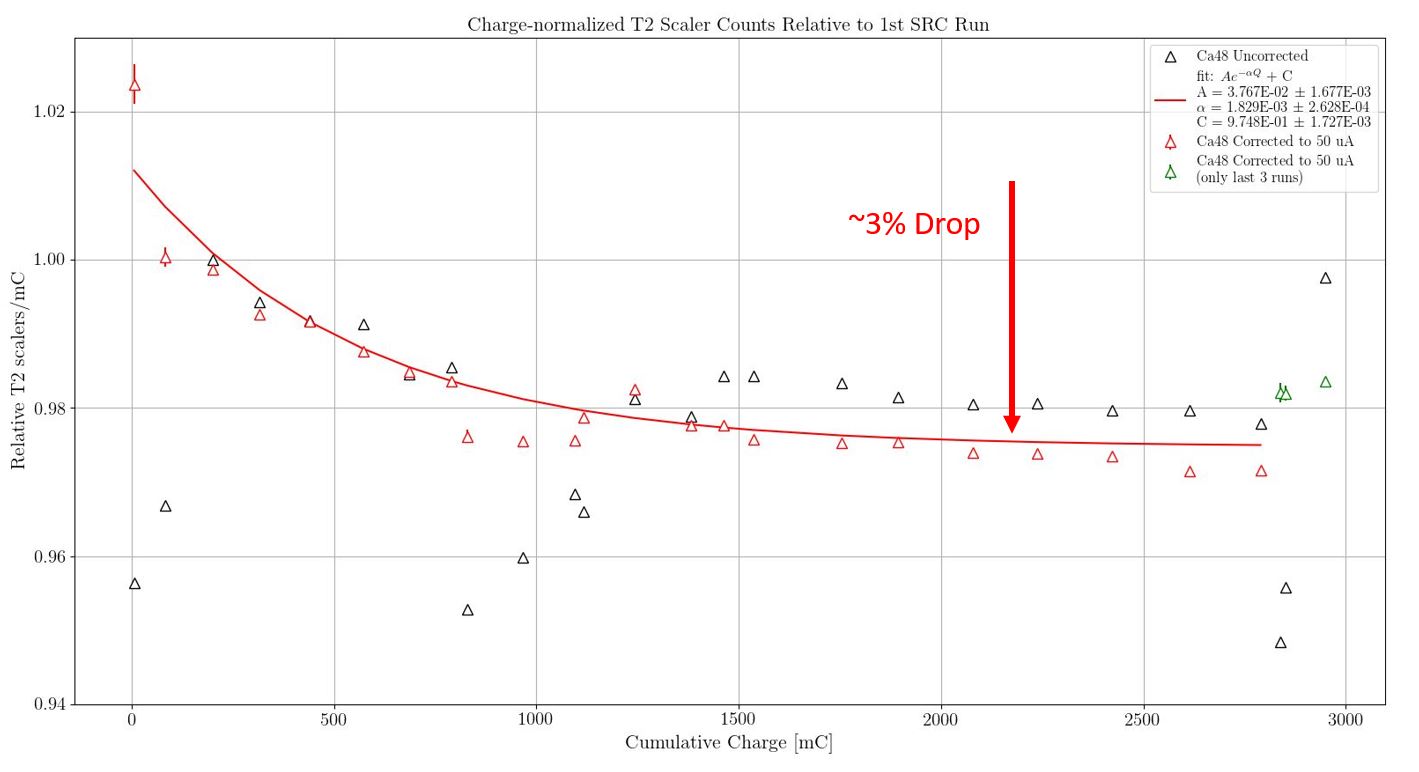}
    \caption{The Ca$(e,e')$ SHMS trigger rate per milliCoulomb as a function of accumulated beam charge.  The red points are corrected for rate effects.  The decreases indicates the evaporation of the oil as the run progressed.  The red line is an exponential fit to the red points.}
    \label{ca48-oil}
\end{figure}

\section{Radiative and transparency corrections}
We calculated the ratio of the radiative corrections ($R_A/R_C$) and transparency corrections ($T_A/T_C$) for different nuclei relative to $^{12}$C and applied these corrections to the measured cross section ratios, see Table~\ref{tab:corr}.

\begin{table}[h!]
    \centering
    \begin{tabular}{|c|c|c|}
     \hline
        Ratio  & $R_A/R_C$  & $T_A/T_C$\\ \hline
        $^{9}$Be/C  & 1.00 $\pm$ 0.02 & 1.17 $\pm$ 0.05 \\ \hline
        $^{10}$B/C  & 1.00 $\pm$ 0.02 & 1.09 $\pm$ 0.02 \\ \hline
        $^{11}$B/C  & 1.00 $\pm$ 0.02 & 1.04 $\pm$ 0.02 \\ \hline
        $^{40}$Ca/C  & 0.97 $\pm$ 0.02 & 0.75 $\pm$ 0.03 \\ \hline
        $^{48}$Ca/C  & 0.97 $\pm$ 0.02 & 0.68 $\pm$ 0.03 \\ \hline
        $^{54}$Fe/C  & 0.99 $\pm$ 0.02 & 0.66 $\pm$ 0.03 \\ \hline
        $^{197}$Au/C  & 0.81 $\pm$ 0.3 & 0.44 $\pm$ 0.02 \\ \hline

    \end{tabular}
    \caption{Ratio of radiative corrections ($R_A/R_C$) and transparency corrections ($T_A/T_C$) and their associated uncertainties.}
    \label{tab:corr}
\end{table}

\section{Cut variation}
The values of the cuts applied to both the kinematic variables and the collimator apertures were randomly sampled from Gaussian distributions with the mean and   corresponding $\sigma$  for each variable  listed in Table~\ref{cutvar}. A total of 10,000 cut sets were generated. For each set, the cross-section ratio was calculated.  The 10,000 resulting ratios yielded an approximately Gaussian distribution. The standard deviation of this distribution was assigned as the systematic uncertainty associated with the cut variation, see Table~\ref{tab:cutsen2}.

\begin{table}[htbp!]
\begin{center}
\renewcommand{\arraystretch}{1.0}
\begin{tabular}{| l | l | l |}\hline
 Cut variables     & Means & $\pm \sigma$\\ \hline\hline
 $P_{miss}^{min}$ (GeV/$c$) & 0.375 & $\pm$ 0.0125\\ \hline
 $P_{miss}^{max}$ (GeV$c$) & 0.7 & $\pm$ 0.05\\ \hline
 $x_{bj}^{min}$ & 1.2 & $\pm$ 0.05\\ \hline
 $\theta_{rq}^{max}$ (deg)& 40 & $\pm$ 2\\ \hline
 $Q^{2, min}$ (GeV$^{2}$/$c^2$) & 1.8 & $\pm$ 0.05 \\ \hline
 HMS Collimator Size &  & $\pm 4$\%\\ \hline
SHMS Collimator Size &  & $\pm 4$\%\\ \hline
\end{tabular}
\end{center}
\caption{The mean values and standard deviations of the Gaussian distributions for the cut variations.}
\label{cutvar}
\end{table}

\section{Summary of Cross section and uncertainties}

Table ~\ref{data-model} lists the per-proton cross-section ratios for different nuclei to C.  These values are plotted in Fig.~3 in the main text.

\begin{table}[ht!]
    \begin{center}
    \begin{tabular}{| l | l | l | l | l | l | l |}
    \hline
    Ratio & $\delta_{rad} (\%) $ & $\delta_{\text{cut}}$ (\%) & $\delta_{trans}$(\%) & $\delta^{\text{total}}_{sys}$ (\%) & $\delta_{\text{stat}} (\%)$ & $\delta^{\text{total}} (\%)$\\ \hline
    
    $^{9}$Be/$^{12}$C  & 2.0 & 1.3 & 4.0 & 4.47 & 2.1 & 5.0 \\ \hline
    $^{10}$B/$^{12}$C & 2.0 & 1.0 & 2.0 & 3.0 & 2.6 & 4.0 \\ \hline
    $^{11}$B/$^{12}$C & 2.0 & 0.9 & 2.0 & 3.0 & 2.5 & 3.9 \\ \hline
    ${^{12}}$C/${^{12}}$C & 0 & 0 & 0 & 0 & 0 & 0 \\ \hline
    $^{40}$Ca/$^{12}$C & 2.0 & 2.2 & 4.0 & 5.0 & 1.9 & 5.3 \\ \hline
    $^{48}$Ca/$^{12}$C & 2.0 & 3.1 & 4.0 & 5.4 & 1.9 & 5.8 \\ \hline
    $^{54}$Fe/$^{12}$C & 2.0 & 2.4 & 4.0 & 5.1 & 2.1 & 5.5 \\ \hline
    $^{197}$Au/$^{12}$C & 4.0 & 4.9 & 4.0 & 7.5 & 2.7 & 8.0 \\ \hline
    \end{tabular}
    \end{center}
    \caption{Experimental uncertainties.  $\delta_{rad}$, $\delta_{cut}$, and $\delta_{trans}$ are the systematic uncertainties associated with the radiative corrections, cut variations, and transparency corrections, $\delta^{\text{total}}_{sys}$ is the total systematic uncertainty, $\delta_{stat}$ is the statistical uncertainty, and $\delta^{total}$ is the combined systematic plus statistical uncertainty, added in quadrature.  All uncertainties are in percent.   }
    \label{tab:cutsen2}
    \end{table}

\begin{table}[ht!]
 \begin{center}
 \small
 \begin{tabular}{| l | c | l | l | l | l | l | l |} \hline
     
     Target
      &  Data  &  AV18  &  SRG  &  LCA  
      & \begin{tabular}[c]{@{}l@{}}$l=0$\\$n=0$\end{tabular}
      & \begin{tabular}[c]{@{}l@{}}$l=0$ \\$L=0$\end{tabular} 
      \\ \hline    
    $^{9}$Be   & 0.80 $\pm$ 0.04 & 0.93 & 1.02 & 1.01 &  0.90 & 1.00 \\ \hline
    $^{10}$B   & 0.92 $\pm$ 0.04 & 0.90 & 0.97 & 0.96 &  0.89   & 0.80 \\ \hline
    $^{11}$B   & 1.00 $\pm$ 0.04 & 1.01 & 1.04 & 1.04 &  0.99   & 1.00 \\ \hline
    $^{12}$C   & 1               & 1    & 1    & 1    &  1    & 1\\ \hline
    $^{40}$Ca  & 0.93 $\pm$ 0.05 & 0.85 & 1.16 & 1.10 &  1.62 & 1.00\\ \hline
    $^{48}$Ca  & 1.02 $\pm$ 0.06 & NA   & 1.35 & 1.24 &  1.83 & 1.0 \\ \hline
    $^{54}$Fe  & 1.17 $\pm$ 0.06 & NA   & 1.18 & 1.14 &  1.71 & 1.10 \\ \hline
    $^{197}$Au & 1.26 $\pm$ 0.10 & NA   & 1.58 & 1.35 &  2.55 & 1.10 \\ \hline
    \end{tabular}
    \end{center}
    \caption{Per proton cross-section ratios for different nuclei to C, $(\sigma_A/Z)/(\sigma_C/6)$, for both data and models.} 
    \label{data-model}
\end{table}